\documentclass[preprint,12pt]{emulateapj}

\shorttitle{The Kiloparsec-Scale Jet of 3C\,345}
\shortauthors{Roberts, Wardle, \& Marchenko}
\usepackage{datetime}
\slugcomment{Submitted to the Astronomical Journal 2012 sep 27}

\begin{document}

\title{The Structure \& Linear Polarization of the Kiloparsec-Scale Jet\\of the Quasar 3C\,345}

\author{David H.\ Roberts,\footnote{Visiting Astronomer, National Radio Astronomy Observatory.} John F.\ C.\ Wardle, \& Valerie V.\ Marchenko}
\affil{Department of Physics MS-057, Brandeis University, Waltham, MA 02454-0911}
\email{roberts@brandeis.edu}

\begin{abstract}

Deep Very Large Array imaging of the quasar 3C\,345 at 4.86 and 8.44~GHz has been used to study the structure and linear polarization of its radio jet on scales ranging from 2 to 30~kpc. There is a 7--8~Jy unresolved core with spectral index $\alpha \simeq -0.24$ ($I_\nu \propto \nu^{\alpha}$). The jet (typical intensity 15~mJy/beam) consists of a $2.5\arcsec$ straight section containing two knots, and two additional non-co-linear knots at the end. The jet's total projected length is about 27~kpc. The spectral index of the jet varies over $-1.1 \lesssim \alpha \lesssim -0.5$. The jet diverges with a semi-opening angle of about $9^\circ$, and is nearly constant in integrated brightness over its length. A faint feature north-east of the core does not appear to be a true counter-jet, but rather an extended lobe of this FR-II radio source seen in projection. The absence of a counter-jet is sufficient to place modest constraints on the speed of the jet on these scales, requiring $\beta \ga 0.5$. Despite the indication of jet precession in the total intensity structure, the polarization images suggest instead a jet re-directed at least twice by collisions with the external medium.

Surprisingly, the electric vector position angles in the main body of the jet are neither longitudinal nor transverse, but make an angle of about $55^\circ$ with the jet axis in the middle while along the edges the vectors are transverse, suggesting a helical magnetic field. There is no significant Faraday rotation in the source, so that is not the cause of the twist. The fractional polarization in the jet averages 25\% and is higher at the edges. In a companion paper \citep{R+W} it is shown that differential Doppler boosting in a diverging relativisitic velocity field can explain the electric vector pattern in the jet.

\end{abstract}


\section{INTRODUCTION}

The compact radio source 3C\,345 (J1642+3948) is a 16th magnitude quasar at a redshift of $z = 0.593$ (6.5 kpc/arcsecond in concordance cosmology, $h = 0.73$).  This is one of the best studied quasars at a wide range of wavelengths and angular resolutions  because it is nearby and bright.\footnote{There are over 1000 references to 3C\,345 in the NASA Extragalactic Database (http://nedwww.ipac.caltech.edu/), to which we refer the reader for further information.} The source is variable in X-ray, optical, and radio bands on time scales ranging from minutes to months, and it is highly polarized at optical and radio wavelengths (\cite{Kinman}, \cite{Aller}). The radio structure was studied by \cite{KWR} on angular scales from $0.3\arcsec$ through $30\arcsec$ (2 to 200~kpc) using multi-array VLA data at 5~GHz. They were able to trace the radio source from the core north-west along the jet, which then turns northward and eventually loops around to the east and then to the south (see also \citet{MBP}). \cite{KWR} concluded that this highly distorted structure was the result of projection, and that 3C\,345 was a Fanaroff-Riley Type~II radio source \citep{FR} seen nearly end-on. 3C\,345 has been observed extensively by Very Long Baseline Interferometry, and is superluminal with inferred Lorentz factor $\Gamma$ up to at least 20 \citep{Lister}.  Thus it is assumed that the jet near the core is aligned within an angle $ i \sim 1/\Gamma$ of the line of sight, consistent with the interpretation of the large-scale structure. Since at larger distances the jet is strongly curving, the inclination of the jet within about $0$-$4\arcsec$ of the core---the region studied below---is not well determined. This uncertainty makes it impossible to de-project the jet with complete confidence; however, it seems certain that bends and opening angles are exaggerated by projection.

In this paper we use high-resolution, high dynamic range VLA images at 5 and 8~GHz to study the structure and linear polarization of 3C\,345. The observations and data reduction are described in \S\ref{s:obs}, the properties of the jet described in \S\ref{s:jets}, we discuss the results in \S\ref{s:Discussion}, and our conclusions are presented in \S\ref{s:conclude}.

\section{OBSERVATIONS AND DATA REDUCTION}
\label{s:obs}

Interferometer data from 1989 January 7--8 (epoch 1989.02)  were obtained from the VLA data archive (experiment code AR196). The array was in the A configuration with 27 working antennas. The frequencies used were 4.86~GHz (C~band, $\lambda6.2\mbox{ cm}$) and 8.44~GHz (X~band, $\lambda3.6\mbox{ cm}$), with a bandwidth of 50~MHz for each of the two IF systems at each band. The data were edited and calibrated in AIPS \citep{AIPS}. The nearby source 1635+381 was used for phase calibration, and the primary flux calibrator 3C\,286 was used for amplitude calibration. To obtain the maximum dynamic range in the images a baseline-dependent calibration was executed with the task BLCAL. The baseline calibration tables were determined from the data for 3C\,286 and its AIPS CLEAN component  models. The tables were then attached to the data files of 3C\,345 for their respective frequency bands. It was found that making a baseline correction using an image of the bright source 3C\,84 made from data collected about 12 hours earlier produced essentially the same results at 5~GHz, but was superior at 8~GHz. Hence this correction was used at the higher frequency. Baseline-corrected images had signal-to-noise about a factor of two better than those made without baseline correction, and were judged to have higher image fidelity.

Polarization calibration was done with the phase calibrator using the task PCAL. The position angle of linear polarization on the sky was set by 3C\,286. We estimate that the absolute position angle calibration is good to within $\pm 3^\circ$ at 5~GHz and $\pm 5^\circ$ at 8~GHz. 

Images were made using AIPS and the Caltech difference imaging package DIFMAP \citep{DIFMAP} as follows. Using DIFMAP, at each step in the SELFCAL--CLEAN cycle, a tight CLEAN box was placed around the brightest remaining flux and purely positive CLEAN components were removed until the process ended. Then a 15~minute phase-only SELFCAL was done, and the boxing-CLEANing-SELFCAL resumed. The $u$-$v$ weighting was changed and $u$-$v$ tapers applied as required to recover more flux. Only when this process terminated was an amplitude SELFCAL used, with a 60~minute timescale. Phase-only SELFCALs were resumed until the process terminated. Finally, the self-calibrated visibilities were moved into AIPS and the final images made with the task IMAGR using the restoring beams listed in Table~\ref{tab:maps}.  To create images that could be compared at the two frequencies, uniform weighting (ROBUST $= -5$) was used for 5~GHz and natural weighting (ROBUST $= 5$) for 8~GHz,  a $u$-$v$ taper with parameters (850, 800) k$\lambda$ was used at both frequencies, and identical circular restoring beams of full-width at half-maximum $0.30\arcsec$ were applied. To study the divergence of the jet, full-resolution images were made from the data at 8~GHz; it was judged that ``optimum weighting'' (ROBUST parameter 0) and a CLEAN beam of $0.20\arcsec$ gave the best compromise between resolution and signal-to-noise. To simplify the analysis and image display, most of the images were rotated by $-50^\circ$ in position angle.

In Table~\ref{tab:maps} we give the parameters of the images and the figures. The root-mean-square (RMS) noise reported is from a $5\arcsec \times 5\arcsec$ box centered $7.5\arcsec$ southeast of the core, and is reasonably representative of the uncertainties in the images (see below).

\section{PROPERTIES OF THE 3C\,345 JET AT 5 AND 8 GHz}
\label{s:jets}

\subsection{Total Intensity Structure}

\begin{figure}[t]{}
\includegraphics[angle=0,width=1.0 \columnwidth]{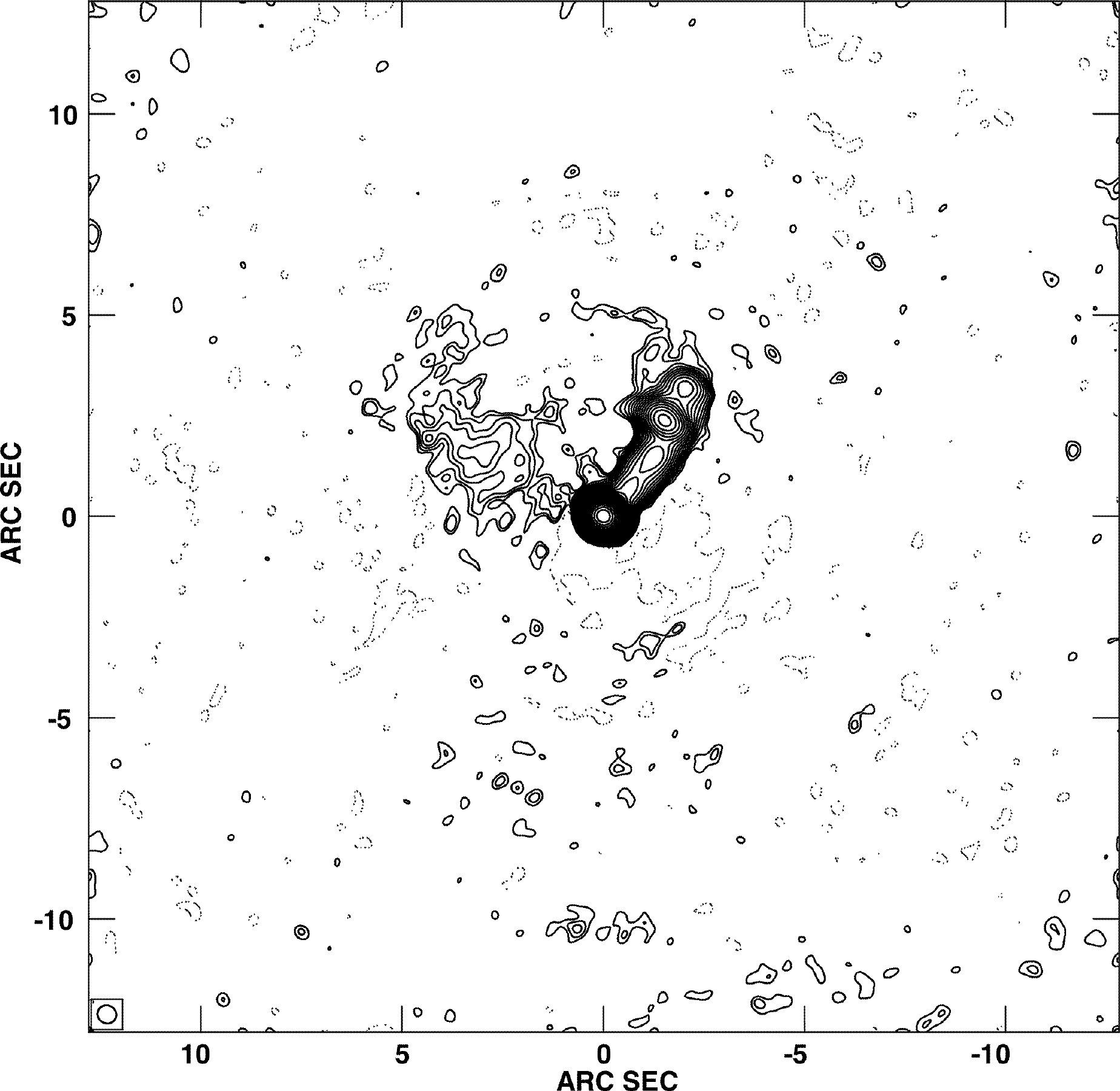}
\caption{Total intensity image of the quasar 3C\,345 made from 5~GHz VLA A-array data, natural weighting (ROBUST = 5), full field showing the orientation of 3C\,345 on the sky. This image is the mean of images made independently by two of the authors from the same data. See Table~\ref{tab:maps} for image data and figure details.  \label{fig:CFULLBL}}
\end{figure}

\begin{figure}[t]{}
\includegraphics[angle=0,width=1.0 \columnwidth]{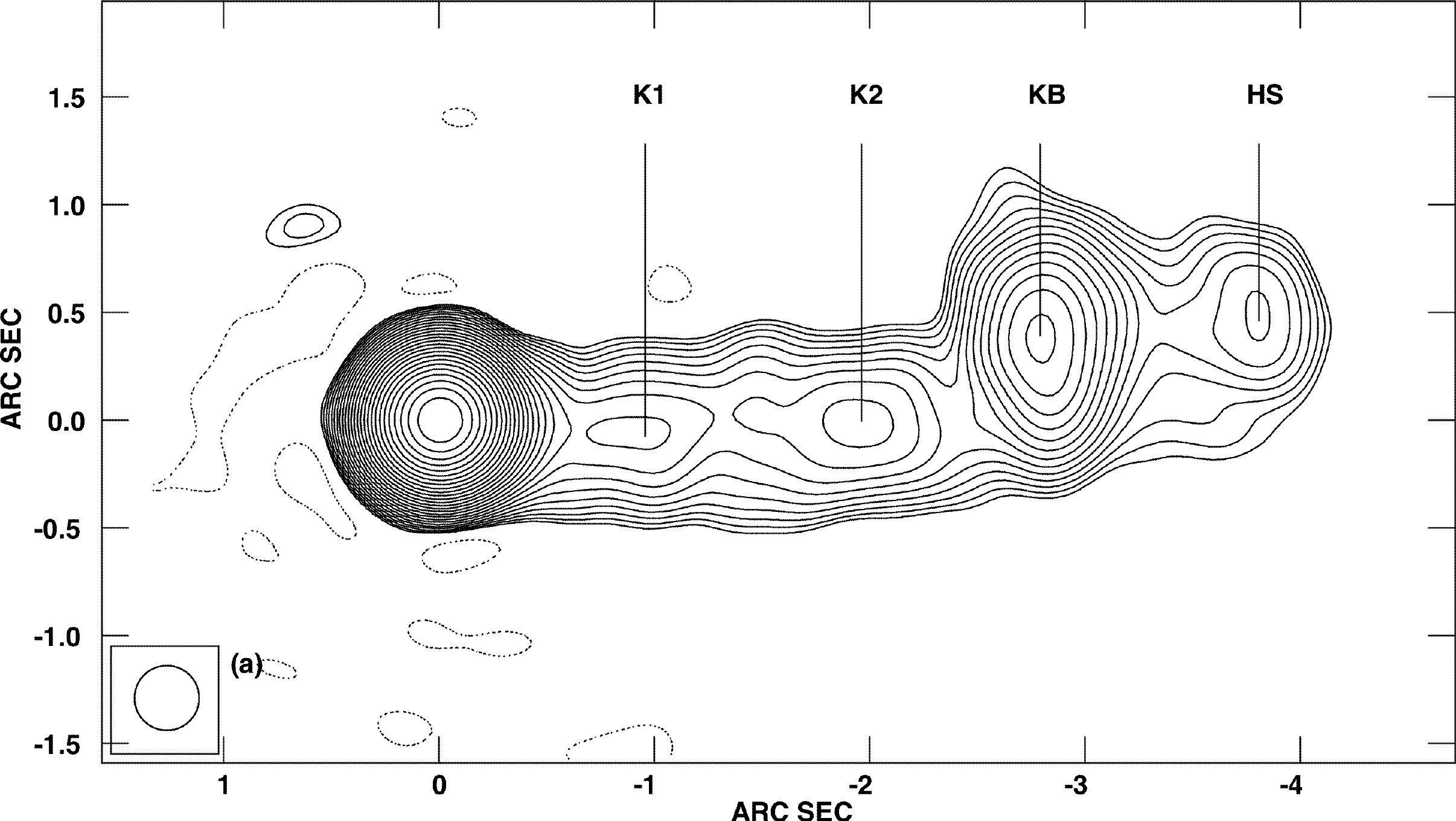}
\includegraphics[angle=0,width=1.0 \columnwidth]{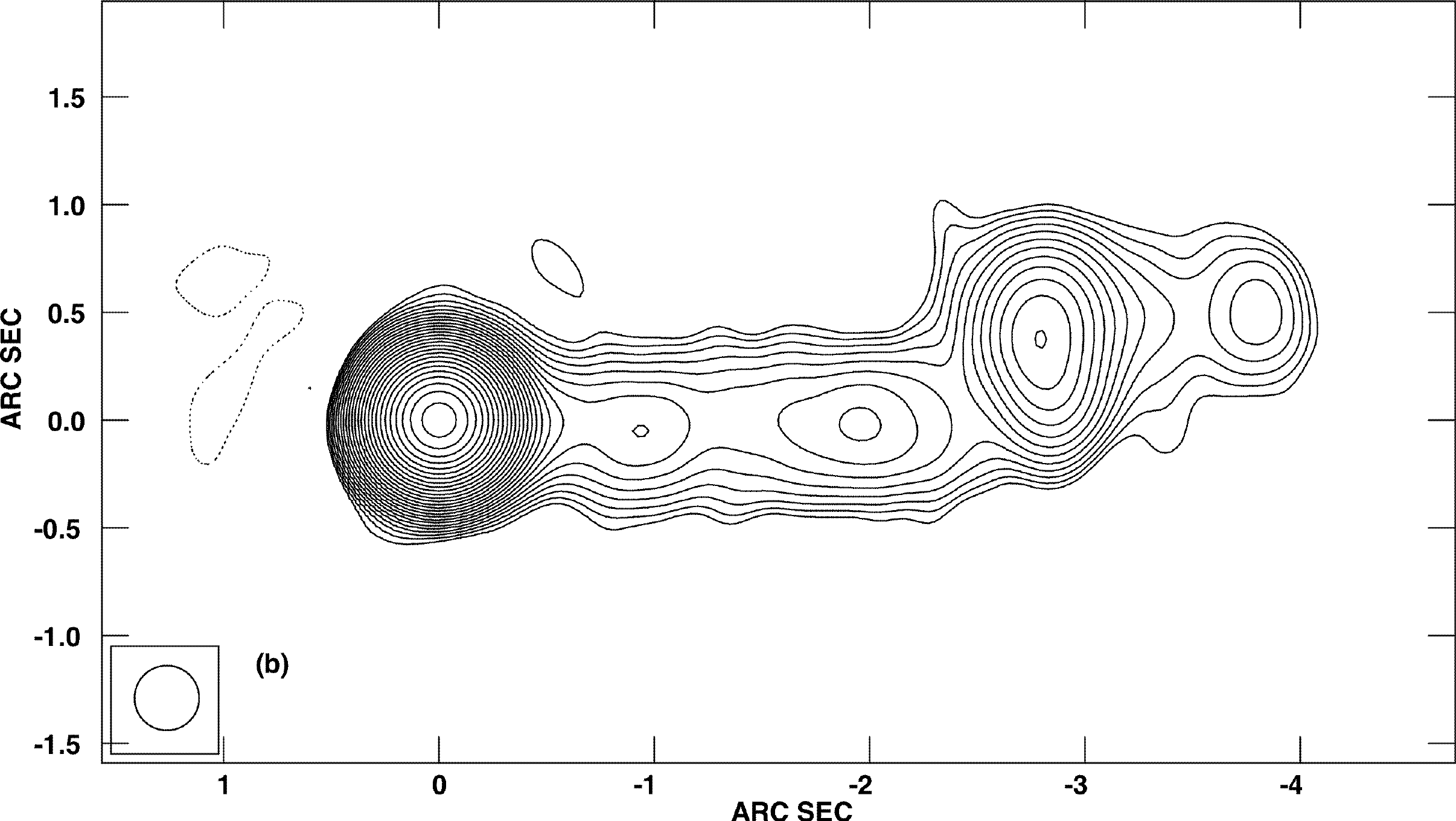}
\caption{Total intensity images of the quasar 3C\,345 made from (a) 5 \& (b) 8~GHz VLA A-array data. The images are the means of those made by two of the authors, and have been rotated by $-50^\circ$ in position angle. See Table~\ref{tab:maps} for image data and figure details.\label{fig:CXIBL}}
\end{figure}

\begin{figure}[t]{}
\includegraphics[angle=0,width=1.0\columnwidth]{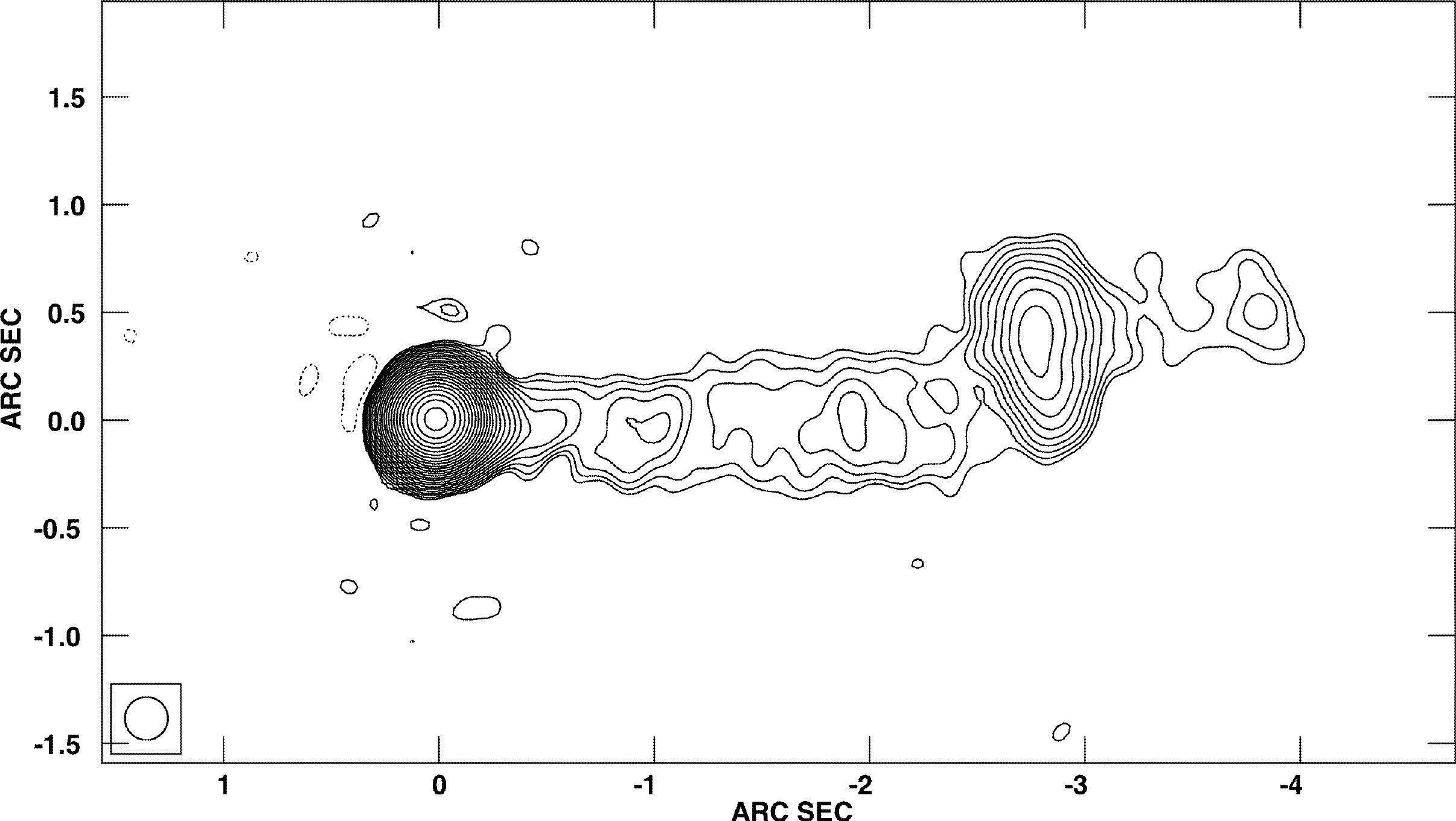}
\caption{VLA image of 3C\,345 at 8~GHz made with ``optimum'' weighting
(ROBUST = $0$). The image is the mean of those made by two of the authors, and has been rotated by $-50^\circ$ in position angle. See Table~\ref{tab:maps} for image data and figure details.\label{fig:HighRes2}}
\end{figure}

A $20\arcsec$-wide field of 3C\,345 at 5~GHz at its correct orientation on the sky is shown  in Figure~\ref{fig:CFULLBL}. The detailed total intensity structures at 5 and 8~GHz are shown in Figure~\ref{fig:CXIBL}.  At the epoch of observation (1989.02), the core of 3C\,345 was essentially unresolved, and had flux densities of  8.1 and 7.1~Jy at 5 and 8~GHz, respectively. Figure~\ref{fig:HighRes2} shows a full-resolution image of 3C\,345 at 8~GHz.

\epsscale{0.65}

\epsscale{0.6}

\epsscale{1.0}

\begin{figure}[t]
\plotone{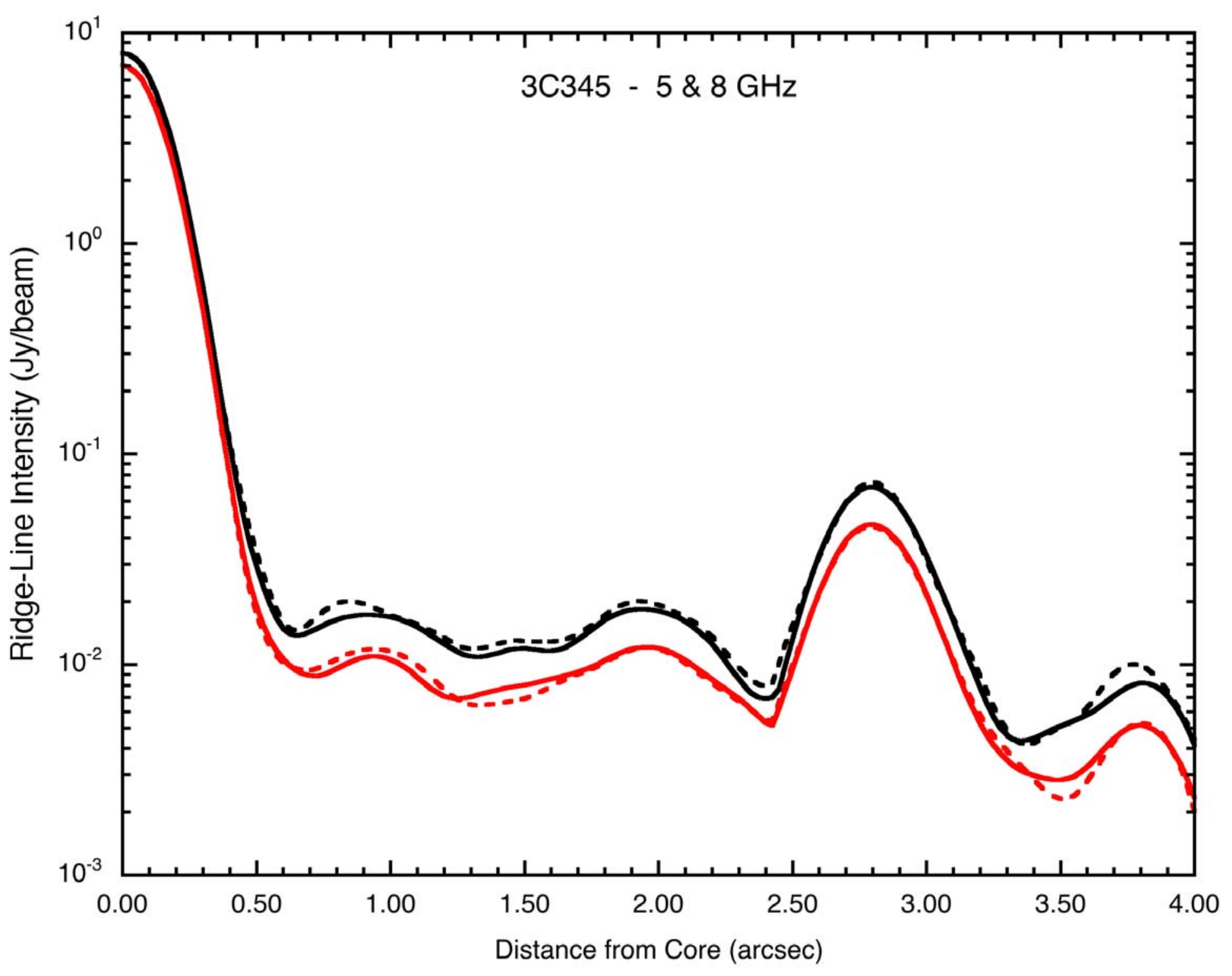}
\caption[]{Comparison of the ridge-line brightness in images made independently by two of the authors from the same 5 and 8~GHz data (Figure~\ref{fig:CXIBL}). The black lines are at 5~GHz, the red lines are at 8~GHz. Note the logarithmic intensity scale. \label{fig:CompareCpk}}
\end{figure}

Since the jet is weak compared to the core, we examined the reliability of the images by comparing those made independently from the same data by two of the authors (DHR \& VVM). The RMS difference between the pairs of images was about 0.5~mJy/beam at each frequency, consistent the RMS noise in the images. Ridge-line intensities at 5 and 8~GHz are compared in Figure~\ref{fig:CompareCpk}, and are consistent with an RMS difference of $\sim 0.5$~mJy/beam. This gives an idea of the dependence of the intensity levels in this faint jet on the personal equation in the hybrid imaging technique. This of course does not mean that the absolute intensity levels are reliable at that level because any calibration uncertainties cancel in this comparison.

\subsection{Total Intensity Properties of the Jet}

The jet consists of a westward straight section\footnote{In those images that have been rotated in position angle by $-50^\circ$ we will refer to directions on the sky as they appear in the rotated images; note that the true position angles are ``West" = $-40^\circ$, ``East" = $+140^\circ$, ``North" = $+50^\circ$, and ``South" = $-130^\circ$.} of approximate projected length $2.25\arcsec$ (16~kpc), with two ``knots'' that we will call ``K1'' and ``K2,'' and two additional bright spots at the west end of the jet that do not lie on the same east-west line. To avoid confusion with K1 and K2, we will refer to the first of these spots as the ``knob,'' denoted ``KB,'' and the final one as the ``hotspot,'' or ``HS;'' see Figure~\ref{fig:CXIBL}a. Examination of the polarization images below suggests that we further subdivide the knob into a northern piece we will refer to as ``KBN'' and a southern piece ``KBS.''

\subsubsection{Spectral Index}

\begin{figure}[t]{}
\includegraphics[angle=0,width=1.0\columnwidth, clip]{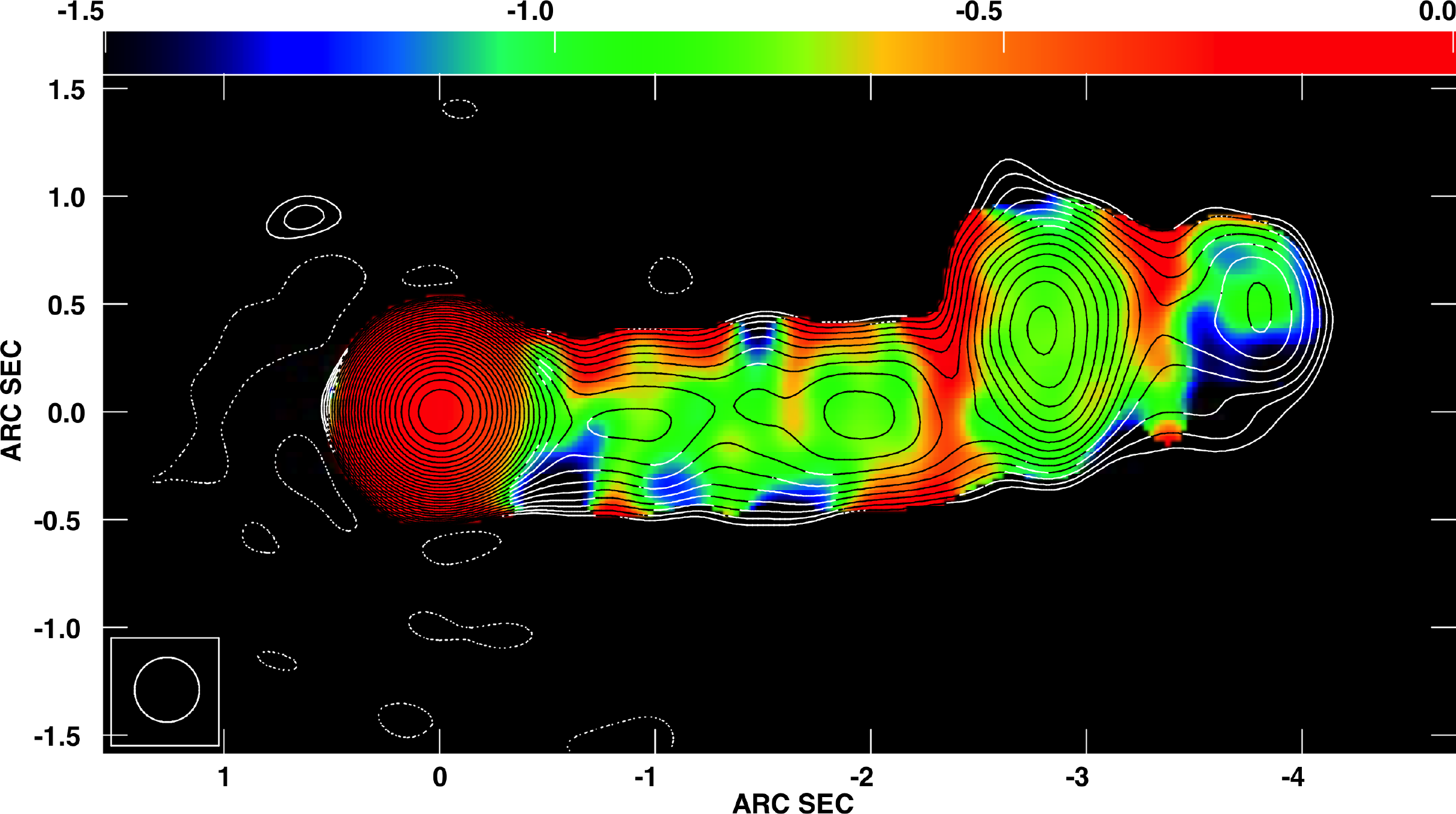}
\caption{Spectral index image of the quasar 3C\,345 made from 5 \& 8~GHz data, shown over contours of total intensity at 5~GHz. The color represents spectral index on the scale to the top. The image was derived from Figure~\ref{fig:CXIBL}, and has been rotated by $-50^\circ$ in position angle. See Table~\ref{tab:maps} for image data and figure details.\label{fig:SpixCX}\\}
\end{figure}

\begin{figure}[t]{}
\plotone{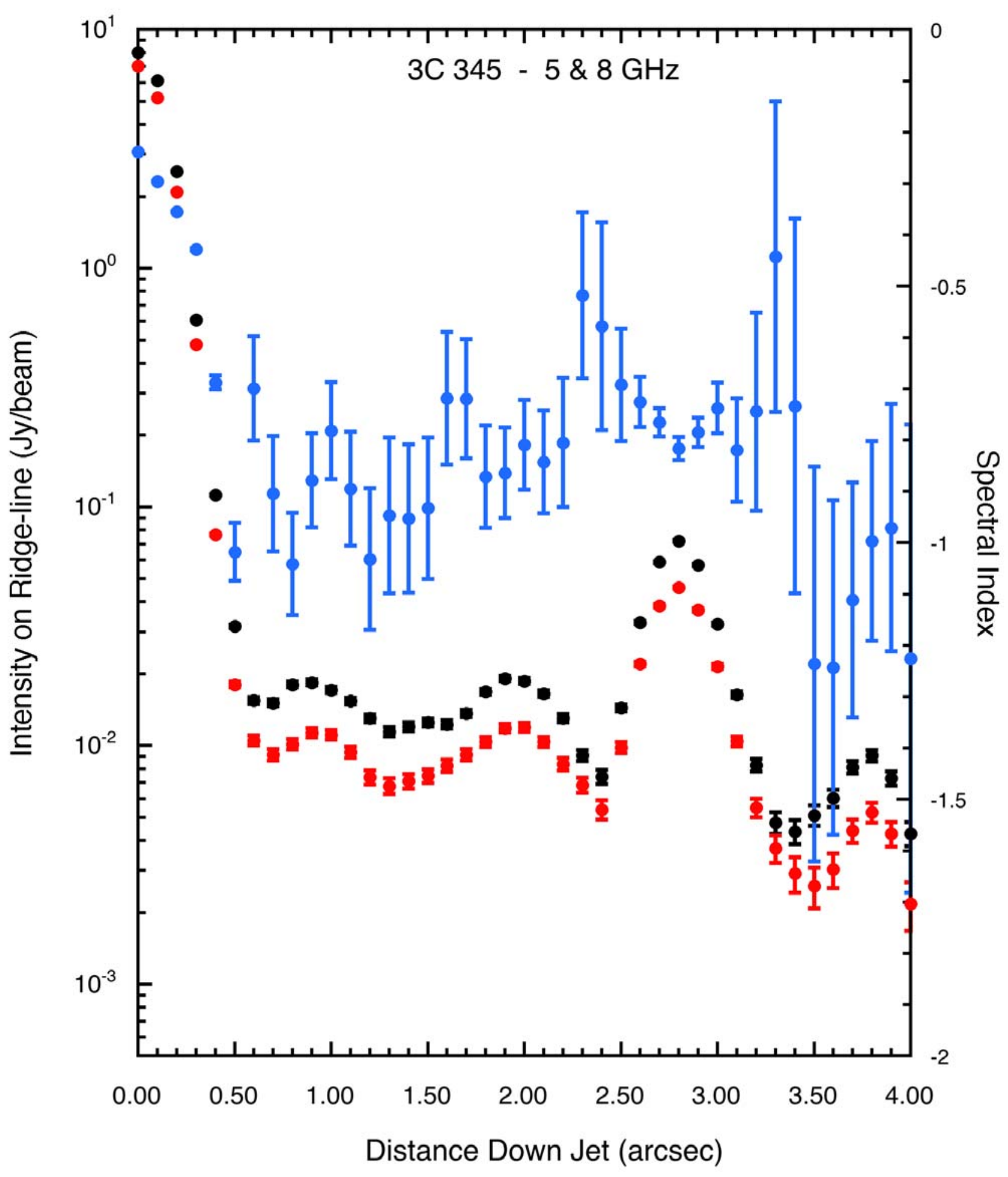}
\caption{Intensity (5~GHz in black, 8~GHz in red) and spectral index (blue) along the ridge-line as a function of distance from the core, derived from the images in Figure~\ref{fig:CXIBL}.  The error bars correspond to an RMS uncertainty of 0.5~mJy/beam. \label{fig:alpha+I}}
\end{figure}

Figure~\ref{fig:SpixCX} shows a spectral index image in color over total intensity contours, made from from the images in Figure~\ref{fig:CXIBL}. The core has a rather flat spectrum with $\alpha = -0.24$, while the spectral index of the jet along the ridge-line ranges over $-1.1 \lesssim \alpha \lesssim -0.5$ $(I_\nu \propto \nu^{\alpha})$. The spectrum is significantly steeper along the south edge of the jet yet flatter on the north side, and flatter between K2 and KB, and between KB and HS. The spectral index is plotted along with the ridge-line intensities in Figure~\ref{fig:alpha+I}. The error bars for $\alpha$ were determined from the estimate of the systematic uncertainties in the intensities derived by comparison of pairs of independent images made from the same data.

\epsscale{1.0}

\epsscale{1.}

\epsscale{1.0}

\subsubsection{Is There a Counter-jet in 3C\,345?}
\label{s:counterjet}

\begin{figure}[t]{}
\plotone{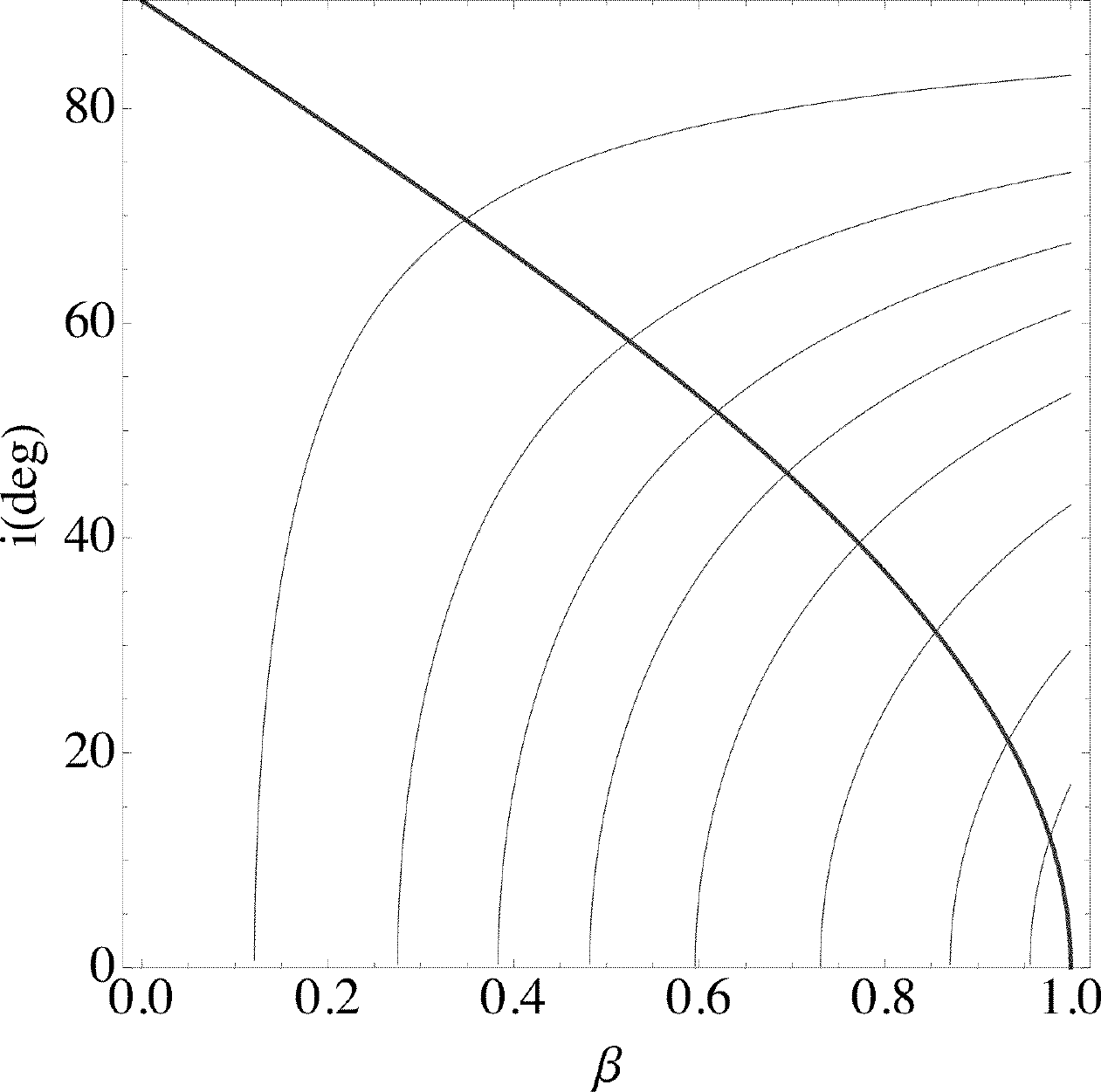}
\caption{Curves of constant jet--counter-jet ratio $J$ in the $(\beta,i)$ plane. From left to right the ratios are 2, 5 10, 20, 50, 200, 2000, \& 5000, computed for a spectral index of $\alpha=-0.85$. The heavy line is the constraint that the photons emerge perpendicular to the jet axis in the frame of the fluid, $\beta = \cos i$.\label{fig:ratios}\\}
\end{figure}

\cite{KWR} detected a possible counter-jet with (true) position angle $\sim55^{\circ}$. In the naturally-weighted image (Figure~\ref{fig:CFULLBL}) we see the same feature in essentially the same location at both 5 and 8~GHz. Is this feature a counter-jet? In our images it is not as ``jet-like'' as in Figure~2 of \cite{KWR}, and it does not meet the criteria of \cite{Perley}. Nonetheless, the feature is real; at a distance of $1\arcsec$ from the core the intensity is about 1~mJy/beam, a factor of about 20 fainter than the main jet at the same core distance, and so we take $J=20$ as a lower limit on the jet--counter-jet ratio. In Figure~\ref{fig:ratios} we show theoretical curves of constant jet-counter-jet ratio in the $(i,\beta)$ plane, assuming that the jet and counter-jet lie along the same line and taking $\alpha = -0.85$. Examination of the curves shows that the fluid speed must be $\beta \ga 0.50$. Thus our images support the conclusion of \cite{KWR} that at least mildly relativistic fluid speeds are required on kiloparsec scales in 3C\,345. More stringent constraints are derived by \citet{R+W} from a model of the polarization. We also show the curve $\beta = \cos i$ that expresses the constraint that the photons we observe are emitted perpendicular to the jet axis in the frame of the fluid \citep{R+W}.

\epsscale{1.0}

\subsubsection{Jet Opening Angle}
\label{s:openingangle}

\begin{figure}[t]{}
\includegraphics[width=0.95\columnwidth]{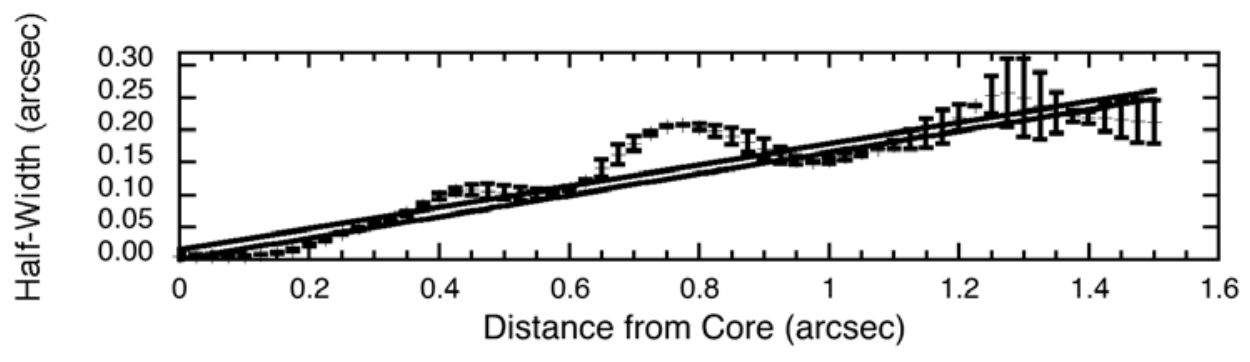}
\caption[]{Deconvolved half-widths and fits at 8~GHz; the data come from Figure~\ref{fig:HighRes2}. The data points are the means of Gaussian fits and simple searches for the half-power points, and the error bars are standard errors using these two estimates. The linear least-squares fits yield half-opening angles of about $9^\circ$.\label{fig:JetWidthsX1}}
\end{figure}

The shape of the jet may play an important role in understanding the peculiar electric vector configuration in this object (see \S\ref{s:twist} below). We examined the divergence of the main body of the jet using a set of transverse slices taken from the image in Figure~\ref{fig:HighRes2}. We estimated the half-width at half-maximum of the jet in two ways, by fitting Gaussian profiles to the slices, and by searching for the location of the half-maxima on each side of the peak, and subtracting the beam quadratically. The results are shown in Figure~\ref{fig:JetWidthsX1}, which shows deconvolved half-widths and fits. The linear least-squares fits yield half-opening angles of $9.31^\circ$ and $9.43^\circ$, depending on whether or not the fit is forced to go through the origin. We adopt an apparent half-opening angle of $\phi_a = 9.4^\circ$.

\epsscale{1.0}

\subsection{Polarization Structure}

\begin{figure}[t]
\includegraphics[angle=0,width=1.0\columnwidth]{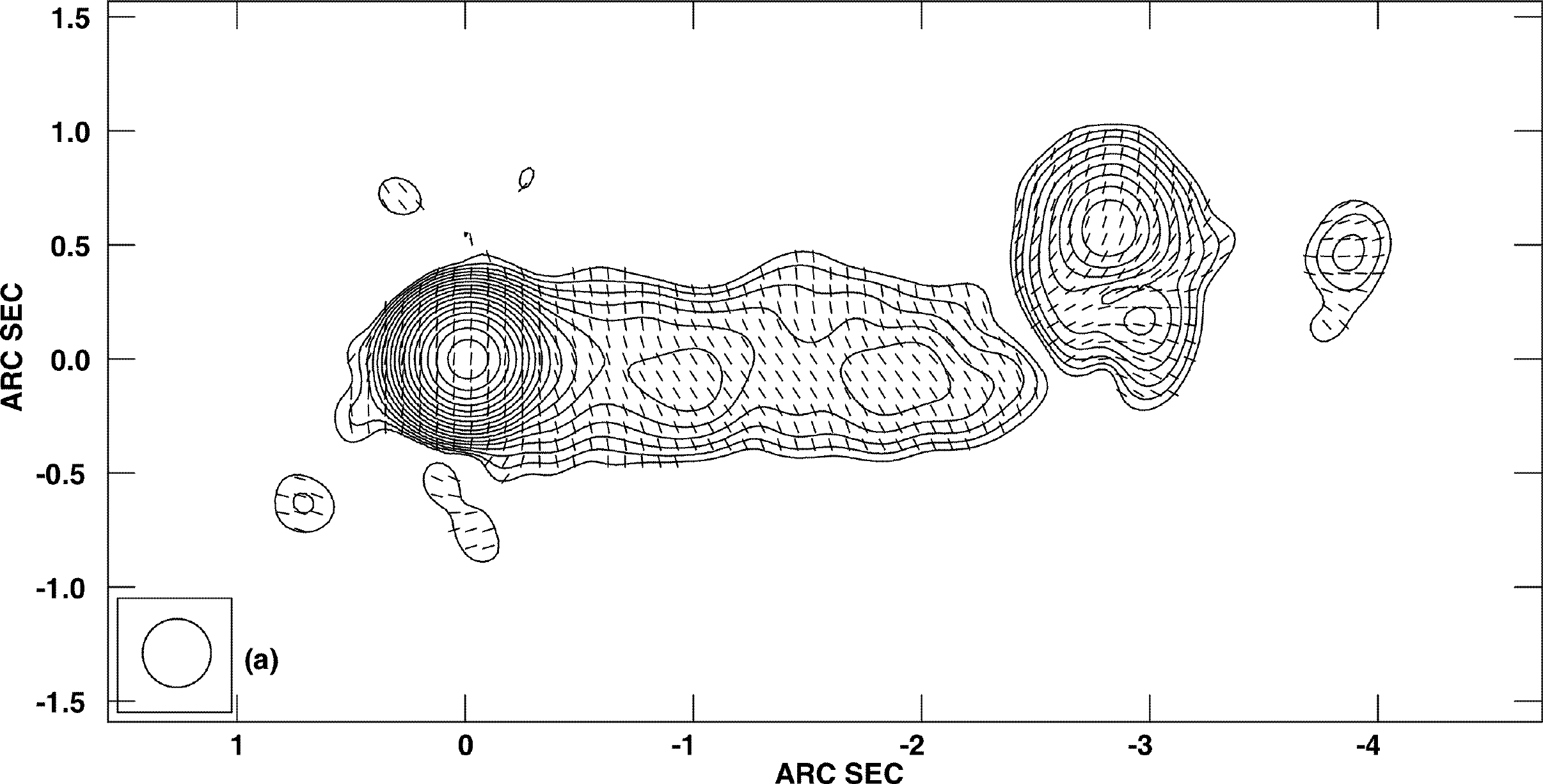}
\includegraphics[angle=0,width=1.0\columnwidth]{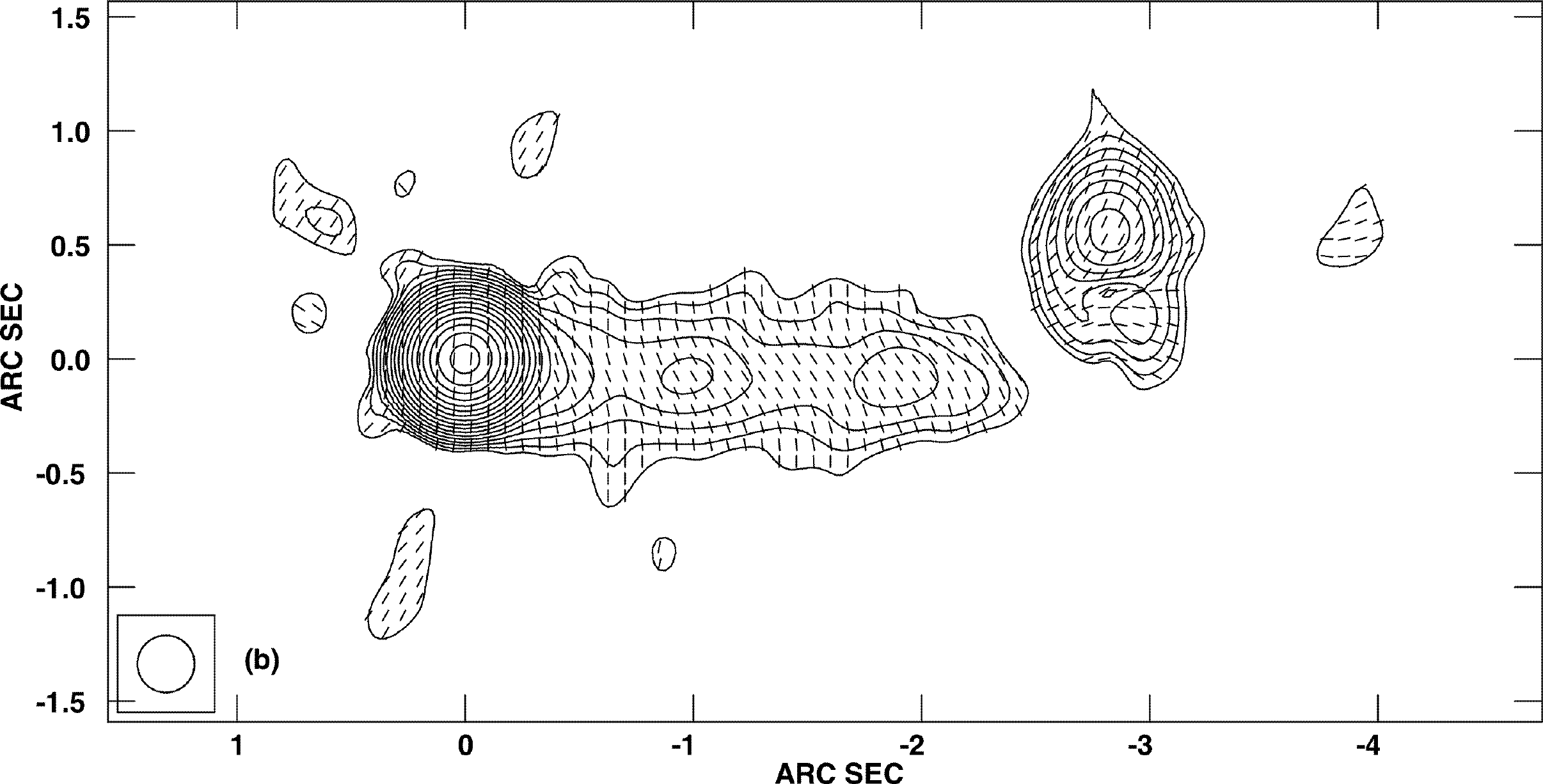}
\caption{Linear polarization images of the quasar 3C\,345 with ticks showing the direction of the observed $\mathbf{E}$ vectors, made from (a) 5 \& (b) 8~GHz VLA A-array data. The contours are of linearly polarized intensity. These images have been rotated by $-50^\circ$ in position angle. See Table~\ref{tab:maps} for image data and figure details.\label{fig:PolnMaps}}
\end{figure}

\begin{figure}[t]
\includegraphics[angle=0,width=1.0\columnwidth, clip]{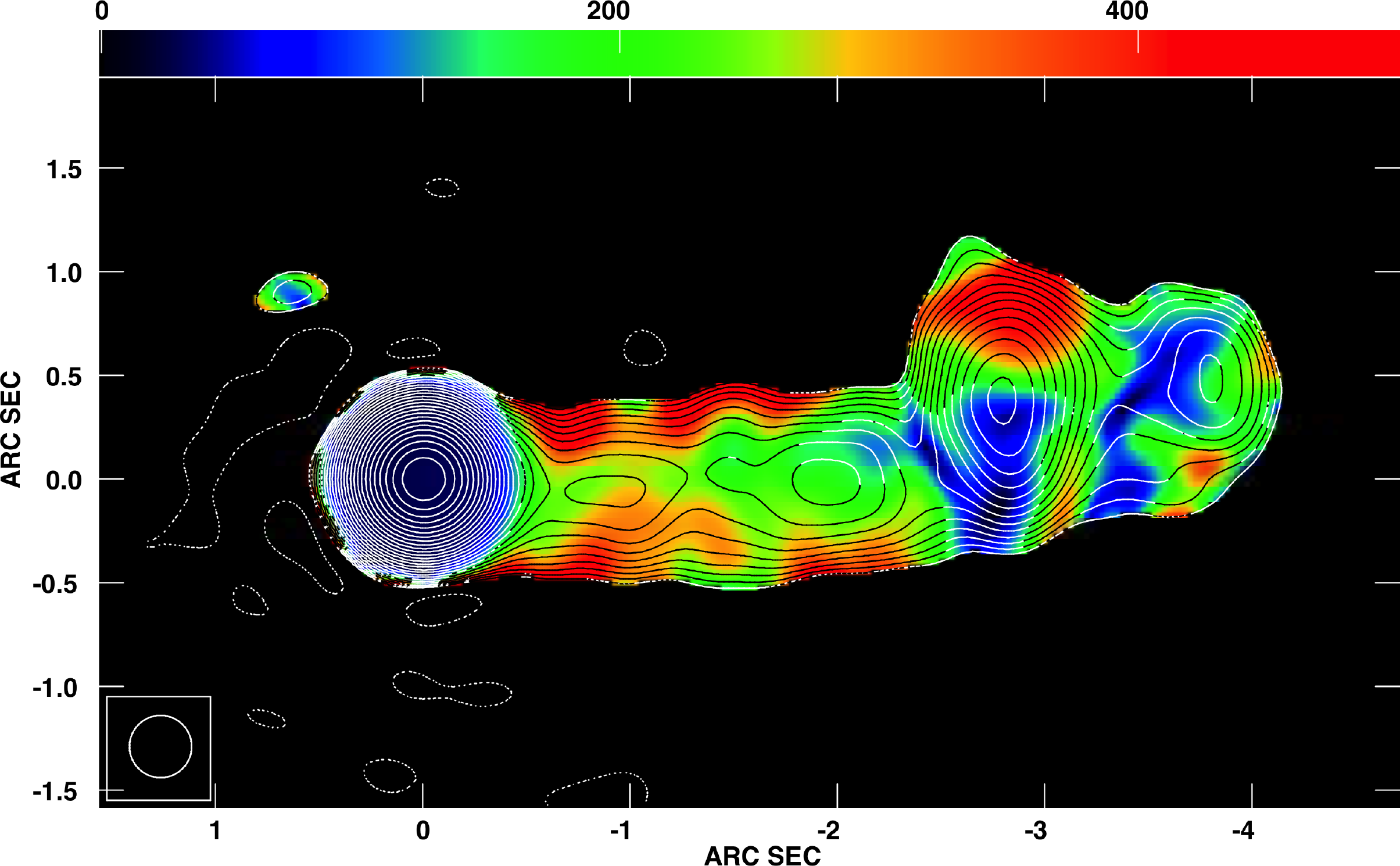}
\includegraphics[angle=0,width=1.0\columnwidth, clip]{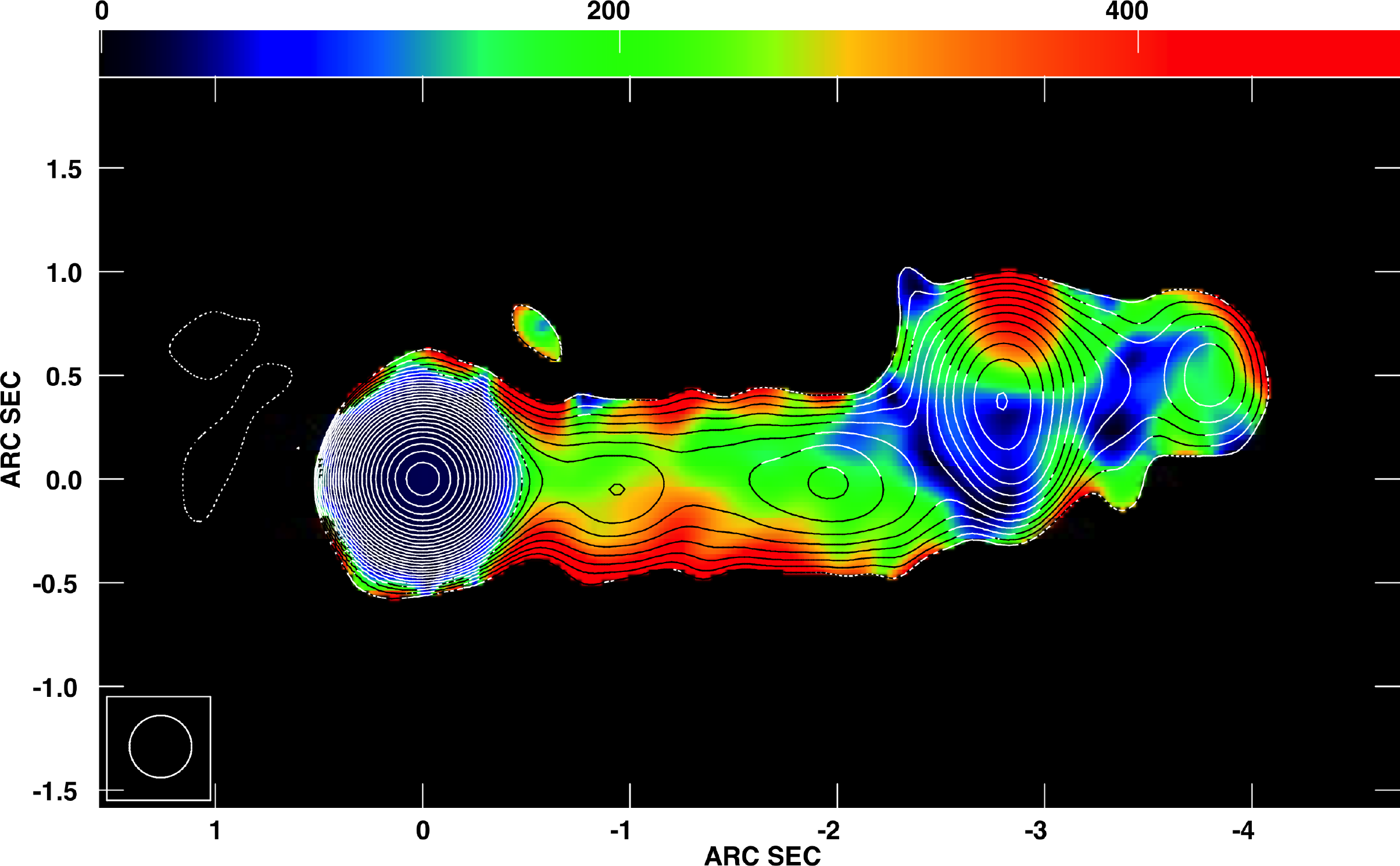}
\caption[]{Color representation of fractional linear polarization over contours of total intensity at (top) 5~GHz and (bottom) 8~GHz. The color scales are $1000m$. See Table~\ref{tab:maps} for image data and figure details.\label{fig:Fpol}}
\end{figure}

The linear polarization\footnote{The notation we use for linear polarization is as follows. The Stokes parameters $Q$ and $U$ are combined to make the complex linear polarization $P = Q + i U = p \, e^{2 i \chi}$, where $m = (Q^2+U^2)^{1/2}/I$ is the fractional polarization, $p = mI$ is the linearly polarized intensity, and $\chi = (1/2) \arctan{(U/Q)}$ is the electric vector position angle.} structures at 5 and 8~GHz are shown in Figure~\ref{fig:PolnMaps}, and the fractional polarization is illustrated in Figure~\ref{fig:Fpol}. We confirm the discovery by \citet{KWR} that the electric vectors in the jet are not transverse to the jet but twisted at a significant angle Here we are able to resolve the polarization of jet in the transverse direction, and find that the twist is greatest, about $35^\circ$, down the center of the jet and vanishes at both edges. The fractional polarization of the jet is about 25\% down the center of the main part of the jet,  greater at the edges, and varies strongly with position in KB and HS.

\begin{figure}[t]{}
\includegraphics[angle=0,width=1.0\columnwidth]{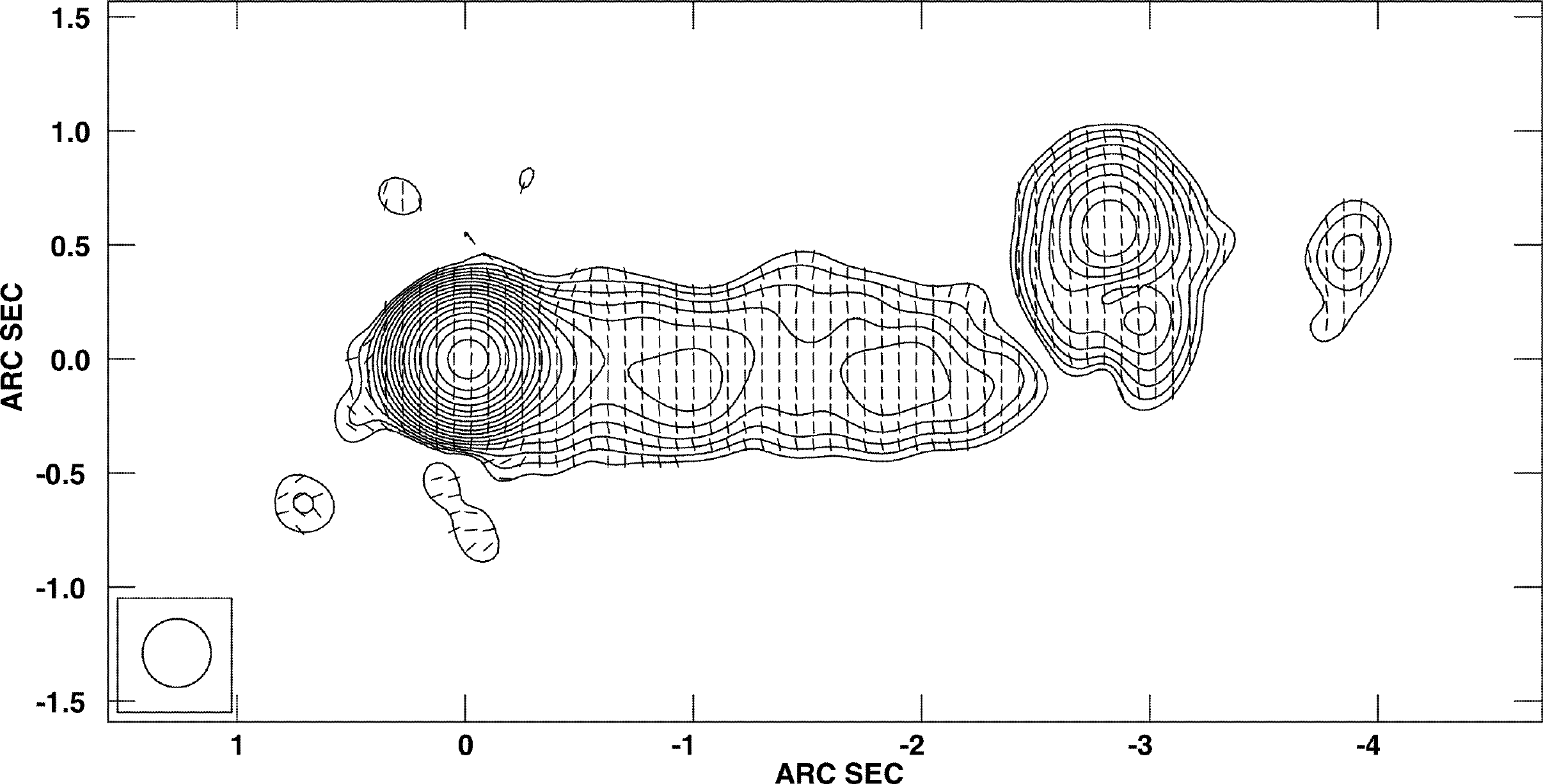}
\caption{Linear polarization position angle differences in 3C\,345 shown over  contours of linear polarized intensity at 5~GHz. Regions with no Faraday rotation will have vertical vectors. This image has been rotated by $-50^\circ$ in position angle. Data from Figure~\ref{fig:PolnMaps}. See Table~\ref{tab:maps} for image data and figure details.\label{fig:DCHI}}
\end{figure}

We first address the possibility that the twist in electric vector position angles is due to Faraday rotation in the jet. This is ruled out as it is apparent from Figure~\ref{fig:PolnMaps} that the electric vector distributions at the two frequencies are nearly identical. Polarization angle differences $\Delta\chi = \chi_5 - \chi_8$ across 3C\,345 are shown in Figure~\ref{fig:DCHI}, where $\Delta\chi$ is displayed as position angle. Any deviations from the vertical indicate Faraday rotation.\footnote{Between 4.86 and 8.44~GHz, $\Delta\chi= 14.6^\circ$ for a rotation measure of $\mbox{RM } = 100\mbox{ rad m}^{-2}$.} Everywhere in the source except the edges, $| \Delta\chi| \lesssim 10^\circ$, corresponding to about $|\mbox{RM}| \lesssim$ 70~rad~m$^{-2}$. The total range of $\Delta\chi$ in the brighter parts, which is independent of absolute polarization position angle calibration, is  $|\Delta\chi_{max} - \Delta\chi_{min}| \lesssim 15^\circ$, or 100~rad~m$^{-2}$. These amounts of Faraday rotation produce only small $\Delta\chi$ at these frequencies, and we conclude that  Faraday rotation is not the cause of the twist in the {\bf E} vector directions in the main body of the jet.

\epsscale{0.7}

\epsscale{0.8}

\epsscale{1.1}

\citet{RJ83} found an integrated rotation measure of $29 \pm 7 \mbox{ rad m}^{-2}$ for 3C\,345, but our measurement at the core yields $-17 \mbox{ rad m}^{-2}$. Our absolute position angle calibration uncertainty of $\pm 5^\circ$ corresponds to a rotation measure uncertainty of about $\pm 30 \mbox{ rad m}^{-2}$, marginally incompatible with the integrated measurement. We have examined data from the University of Michigan Radio Astronomy Observatory for this epoch, and they are compatible with our measurement of negligible Faraday rotation. The source is highly variable and this is the likely cause of the disagreement with the \cite{RJ83} measurement. 

In highly-luminous radio sources with FR~II structure such as 3C\,345 the inferred magnetic field in the jets is typically longitudinal \citep{Bridle94}, as evidenced by orthogonal {\bf E} vectors, while in low-luminosity FR~I sources the inferred field is transverse \citep{Perley}. Thus the linear polarization distribution in the straight portion of the 3C\,345 jet is unusual in that the {\bf E} vectors are neither longitudinal nor transverse to the jet axis.

\section{DISCUSSION}
\label{s:Discussion}

\subsection{Does the Jet in 3C\,345 Precess?}
\label{s:precess}

The non-co-linear $I$ structure of the jet seems to suggest that the jet was pointed in different directions at different times in the past \citep{KWR}. But the polarization structure strongly suggests that this is not the case, and instead that KB contains two impact points of a jet that is twice re-directed by interactions with the external medium, and that comes to an apparent end at the hotspot. We interpret the electric vector configurations at KBS, KBN, and HS as indicating compressed sheets of magnetic field seen roughly edge-on \citep{L80}. In KBS and KBN the jet is redirected by interaction with the ambient medium, and at HS the jet ends in a hot-spot. This interpretation is consistent with a jet inclined at a small angle to the line of sight that is seen roughly from the side due to aberration because its fluid speed is highly relativistic. 

\subsection{Origin of the ``Twist'' in the 3C\,345 Electric Vectors}
\label{s:twist}

Is the magnetic field in the 3C\,345 helical, as it appears to be? There are few similar examples in the literature, the best observed one being M\,87. There the well-resolved jet shows a twisted filamentary structure in the radio \citep{Owen} that has been ascribed to a Kelvin-Helmholtz instability \citep{Hardee}, and the magnetic field structure seems to follow the filaments. Optically it looks quite similar to the radio image \citep{Fraix}. In 3C\,345 we see no filamentary structure in the jet, and suggest instead that the twisted electric field pattern is the result of relativistic transfer effects. A helical magnetic field in a transparent homogeneous cylindrical jet would appear either transverse or longitudinal, not helical, due to cancellations between the back and the front (in a horizontal jet this means that Stokes $U$ is identically zero). Two possibilities for producing a twist in the electric vector pattern suggest themselves; either the symmetry through the jet is broken by opacity or it is broken by relativistic effects. In the latter case, the gentle divergence in the profile of the jet demonstrated in \S\ref{s:openingangle} suggests a similarly diverging velocity field. In such a velocity field, differential Doppler boosting due to differing line-of-sight velocity components in the front and back of the jet will break the symmetry, permitting there to be non-zero $U$, and thus creating a twist to the observed electric vectors if the underlying magnetic field is helical. It is shown in \citet{R+W} that simple models for the magnetic field in 3C\,345 coupled with differential Doppler boosting in a gently diverging jet can produce the ``twist'' seen in the electric vectors in the main body of the jet. This model can be used to place limits on the speed of the jet.

\section{CONCLUSIONS}
\label{s:conclude}

We have made sensitive VLA Array A-array images of the nearby bright quasar 3C\, 345 in total intensity and linear polarization at 5 and 8~GHz. The principle results of the analysis of these images are the following.

\begin{enumerate}

\item 3C\,345 has a bright 7-8~Jy unresolved core with spectral index $\alpha \simeq -0.24$ ($I_\nu \propto \nu^{\alpha}$). 

\item The jet consists of a straight section of projected length about $2.3\arcsec$ (15~kpc), and two bright knots at the end of the jet that do not lie on the same line.

\item The spectral index of the jet ranges over $-1.1 \lesssim \alpha \lesssim -0.5$. The spectrum flattens significantly between the end of the main body of the jet and the first knot at the end, and between that knot and the final one.

\item The jet diverges slightly with an apparent half-opening angle of about $9^\circ$. De-projected, the jet opening angle is significantly smaller \citep{R+W}.

\item There is a faint near-core feature whose base extend from the core at (true) position angle $\sim 55^\circ$. This feature does not appear to be a true counter-jet, but instead is part of an extended lobe seen in projection. The lack of a true counter-jet requires fluid speeds $\beta \ga 0.5$.

\item The fractional polarization in the jet is high, ranging over $0.2 \lesssim m \lesssim 0.5$, and is systematically greater at the edges.

\item The non-co-linearity of the main body of the jet and the knots at its end seems to indicate that the jet was pointed in different directions at different times in the past. However, the polarization structure suggests instead that the knots represent three impact points with the external medium of a jet that is twice re-directed by interactions. 

\item The polarization behavior in the knots at the apparent end of the jet is consistent with the possibility that the jet, despite being at a small angle to the line of sight, is seen effectively from the side due to relativistic effects for a fast-moving flow.

\item Unusually, the electric vector directions in the main body of the jet are neither longitudinal nor transverse, but are twisted. 

\item The ``twist'' in the electric vectors is not due to Faraday rotation, which is a few degrees or less everywhere, and is significantly less in the core, the brightest parts of the jet, and the knots at the end.
 
\end{enumerate}

\section{ACKNOWLEDGMENTS}

The National Radio Astronomy Observatory is a facility of the National Science Foundation, operated under cooperative agreement by Associated Universities, Inc. We thank Dr.~Margo Aller for providing us with the University of Michigan Radio Astronomy Observatory data on 3C\,345. The UMRAO is supported by the National Science Foundation and the University of Michigan. D.H.R. gratefully acknowledges the support of the William R.\ Kenan, Jr.\ Charitable Trust and of the National Radio Astronomy Observatory. J.F.C.W.\ is supported by NSF grant AST-1009262. Any opinions, findings, and conclusions or recommendations expressed in this material are those of the authors and do not necessarily reflect  the views of the National Science Foundation. 

Facilities: \facility{VLA (A array, data archive (experiment code AR196))}.

\clearpage

\begin{deluxetable}{cccccc}
\tabletypesize{\scriptsize}
\tablecaption{Image \& Figure Properties for 3C\,345.}
\tablewidth{0pt}
\tablecolumns{5}
\tablehead{\colhead{Figure} & \colhead{RMS} 
& \colhead{Beam FWHM} & \colhead{Peak} & \colhead{Lowest Contour} \\
\colhead{} & \colhead{(mJy/beam) } 
& \colhead{(arcsec)} & \colhead{(Jy/beam)} & \colhead{(mJy/beam)} \label{tab:maps}}
\startdata
\ref{fig:CFULLBL} &   0.29 &  0.36 & 8.08 & 0.5 \\
\ref{fig:CXIBL}a &  0.46 & 0.30 & 8.14 & 1.0 \\
\ref{fig:CXIBL}b &  0.40 & 0.30 & 7.11 & 1.0 \\
\ref{fig:HighRes2} &  1.4 &0.20 & 6.99 & 1.4 \\
\ref{fig:SpixCX} &  0.46 & 0.30 & 8.14 & 1.0 \\
\ref{fig:PolnMaps}a & 0.78 & 0.30 & 0.227 & 0.71 \\
\ref{fig:PolnMaps}b & 0.99 & 0.30 & 0.209 & 0.71\\
\ref{fig:Fpol}a & 0.78 & 0.30 & 0.227 & 0.71 \\
\ref{fig:Fpol}b & 0.99 & 0.30 & 0.209 & 0.71 \\\
\ref{fig:DCHI} & 0.78 & 0.30 & 0.227 & 0.71 \\ 
\enddata
\tablenotetext{a}{NOTE: All contours are in steps of factors of $\sqrt{2}$.}
\end{deluxetable}


\begin{thebibliography}{} 
\bibitem[Aller(1970)]{Aller} Aller, H.\ D. 1970, \nat, 225, 440.
\bibitem[Bridle et al.(1994)]{Bridle94} Bridle, A.\ H., Hough, D.\ H., Lonsdale, C.\ L., Burns, J.\ O., \& Laing, R.\ A.\ 1994, \aj, 108, 766
\bibitem[Bridle \& Perley(1984)]{Perley} Bridle, A.\ H., \& Perley, R.\ A. 1984, \araa, 22, 319
\bibitem[Fanaroff \& Riley(1974)]{FR} Fanaroff, B.\ L., \& Riley, J.\ M. 1974, \mnras, 167, 31P
\bibitem[Fraix-Burnet, Le Borgne, \& Nieto(1989)]{Fraix} Fraix-Burnet, D., Le Borgne, J.-F. \& Nieto, J.-L. 1989, A\&A, 224, 17.
\bibitem[Hardee \& Eilek(2011)]{Hardee} Hardee, P.\ E. \& Eilek, J.\ A. 2011, \apj, 735, 61
\bibitem[Kinman et. al(1968)]{Kinman} Kinman, T.\ D., Lamia, C., Ciurla, T., Harlan, E., \& Wirtanen, C.\ A. 1968, \apj, 152, 357
\bibitem[Kollgaard, Wardle, \& Roberts(1989)]{KWR} Kollgaard, R.\ I., Wardle, J.\ F.\ C., \& Roberts, D.\ H.\ 1989, \aj, 97, 1550
\bibitem[Laing(1980)]{L80} Laing, R. A. 1980, \mnras, 193, 439
\bibitem[Lister et al.(2009)]{Lister} Lister, M.\ L.\ et al. 2009, \aj, 138, 1874
\bibitem[Murphy, Browne, \& Perley(1993)]{MBP} Murphy, D.\ W., Browne, I.\ W.\ A., \& Perley, R.\ A.\ 1993, \mnras, 264, 298
\bibitem[NRAO(2011)]{AIPS} NRAO 2011, Astronomical Image Processing System, available from the National Radio Astronomical Observatory, Charlottesville, VA, version 31DEC2011
\bibitem[Owen, Hardee, \& Cornwell(1989)]{Owen} Owen, F.\ N., Hardee, P.\ E. \& Cornwell, T.\ J. 1989, \apj, 340, 698
\bibitem[Roberts \& Wardle(2012)]{R+W} Roberts, D.~H., \& Wardle, J.\ F.\ C.\  2012, \apjl, in press (arXiv:1210.2128)
\bibitem[Rudnick \& Jones(1983)]{RJ83} Rudnick, L., \& Jones, T.\ W. 1983, \aj, 88, 518
\bibitem[Shepherd, Pearson, \& Taylor(1994)]{DIFMAP} Shepherd, M.\ C., Pearson, T.\ J., \& Taylor, G.\ B. 1994, \baas, 26, 987
\end{thebibliography}
\end{document}